\documentclass[letterpaper, 10 pt, conference]{ieeeconf}  

\IEEEoverridecommandlockouts                              

\usepackage{balance}       
\usepackage{graphics}      
\usepackage{color}
\usepackage{booktabs}

\usepackage{graphicx}

\usepackage{adjustbox}
\usepackage{caption,subcaption,graphicx}
\usepackage{comment}

\usepackage{ccicons}          

\title{\LARGE \bf
Designing Personalized Interaction of a Socially Assistive Robot for Stroke Rehabilitation Therapy
}

\author{Min Hun Lee$^{1}$, Daniel P. Siewiorek$^{1}$, Asim Smailagic$^{1}$, Alexandre Bernardino$^{2}$ and Sergi Berm{\'u}dez i Badia$^{3}$ \thanks{$^{1}$Carnegie Mellon University, Pittsburgh, PA 15213, USA {\tt\small\{minhunl,dps,asim\}@cs.cmu.edu}}%
\thanks{$^{2}$Instituto Superior T{\'e}cnico, Universidade de Lisboa, Lisbon, Portugal {\tt\small alex@isr.tecnico.ulisboa.pt}}%
\thanks{$^{3}$Madeira Interactive Technology Institute, University of Madeira, NOVA-LINCS, Funchal, Portugal {\tt\small sergi.bermudez@m-iti.org}.}%
\thanks{This work is partially supported by the FCT (SFRH/BD/113694/2015, LARSyS - Plurianual funding 2020-2023), the CMU-PT IntelligentCare project (LISBOA-01-0247-FEDER-045948), and the National Science Foundation (NSF) under grant number CNS-1518865.}
}

\begin{document}

\maketitle
\thispagestyle{empty}
\pagestyle{empty}

\begin{abstract}
The research of a socially assistive robot has a potential to augment and assist physical therapy sessions for patients with neurological and musculoskeletal problems (e.g. stroke). During a physical therapy session, generating personalized feedback is critical to improve patient's engagement. However, prior work on socially assistive robotics for physical therapy has mainly utilized pre-defined corrective feedback even if patients have various physical and functional abilities. This paper presents an interactive approach of a socially assistive robot that can dynamically select kinematic features of assessment on individual patient's exercises to predict the quality of motion and provide patient-specific corrective feedback for personalized interaction of a robot exercise coach. 
\end{abstract}


\section{Introduction}
An early and extensive physical therapy session is an effective intervention for patients with neurological and musculoskeletal problems (e.g stroke)  to regain their functional ability.
However, patients can receive only a limited amount of sessions due to the costs and the shortage of therapists.

Researchers have explored the possibility of supplementing health services with advanced computing and a socially assistive robot \cite{mataric2007socially}. For instance, researchers envision that a socially assistive robot can be integrated into the rehabilitation process by automatically monitoring patient's exercises and providing motivational feedback until the patient's next visits to a therapist \cite{mataric2007socially}. Prior work on robotic exercise coaching systems demonstrates elderly or post-stroke subjects can successfully exercise and stay engaged with a robot over a single \cite{swift2015effects} or multiple sessions \cite{fasola2013socially,gorer2017autonomous}. 
However, in spite of this potential of a robot to monitor and guide exercises, prior work is limited to provide pre-defined corrective feedback on patient's exercise performance (e.g. angular difference with a motion template \cite{fasola2013socially,gorer2017autonomous}).
Generating personalized interaction and feedback for an individual patient still remain a challenge \cite{gorer2017autonomous}.

\begin{figure*}[tp!]
\centering
\begin{subfigure}[t]{.4\textwidth}
\centering
  \includegraphics[width=1.0\columnwidth]{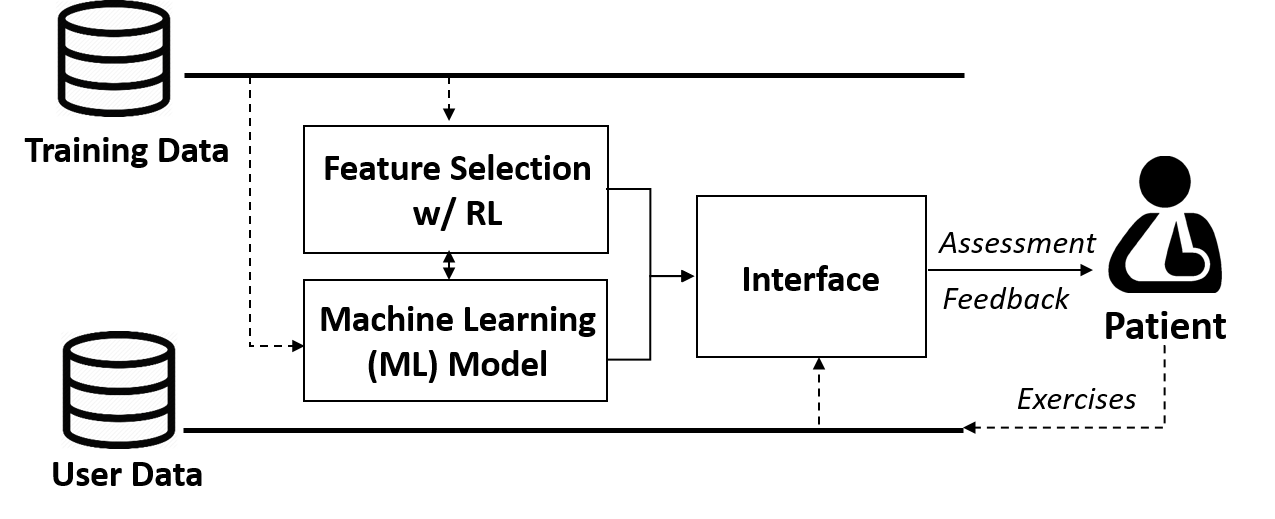}
  \caption{}
  \label{fig:flow-diagram}
\end{subfigure}
\begin{subfigure}[t]{.2\textwidth}
  \centering
  \includegraphics[height=0.91\columnwidth]{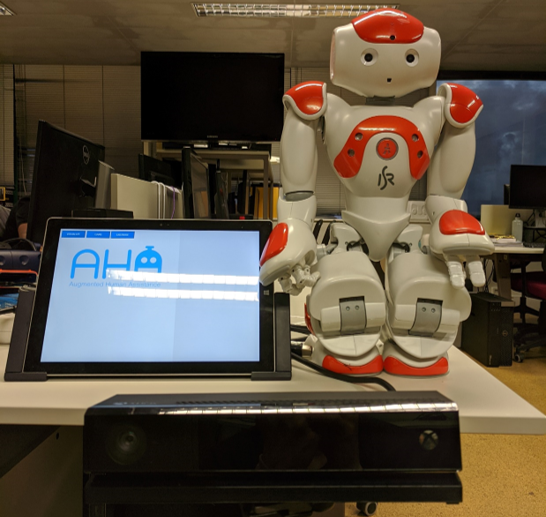}
  \caption{}
  \label{fig:setup}
\end{subfigure}
\begin{subfigure}[t]{.38\textwidth}
  \centering
  \includegraphics[width=1.0\columnwidth]{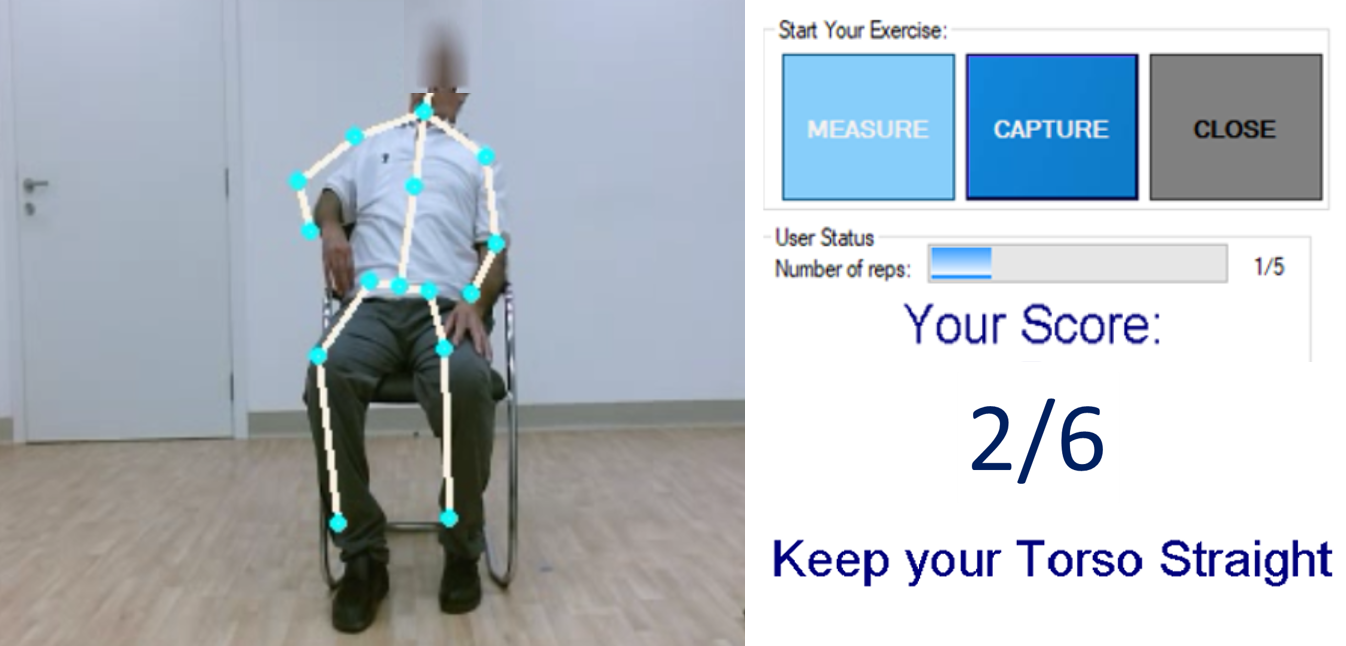}
  \caption{}
  \label{fig:interface}
\end{subfigure}
    \caption{(a) Flow diagram of an interactive approach of an assistive robot for personalized physical therapy. (b) the setup of a system with a Kinect sensor, a tablet with the visualization interface, and the NAO robot. (c) An example output of the visualization interface that presents predicted assessment on patient's exercise performance with corrective feedback.}
\end{figure*}

To address this challenge, this paper presents an interactive approach of a socially assistive robot for personalized post-stroke therapy (Figure \ref{fig:approach})   \cite{ijcai2019-lee}. This approach utilizes reinforcement learning to dynamically select the most important kinematic features of stroke rehabilitation assessment for individual patient's exercise motions to predict the quality of motion \cite{lee2020iha}. Utilizing selected features and patient's held-out normal motions, a robotic system can analyze which features of an affected motion have been deviated from those of normal motions, and generate personalized corrective feedback on a patient's exercise motion (Figure \ref{fig:interface}) \cite{ijcai2019-lee}. 



 \label{fig:approach}
\section{Interactive Approach of a Socially Assistive Robot for Personalized Physical Therapy}
This work aims to support transparent and personalized interaction of a robot exercise coach that utilizes a sparse machine learning model with feature selection \cite{biran2017explanation} to predict the quality of motion and generate corrective feedback (Figure \ref{fig:interface}) with held-out patient's normal motions \cite{ijcai2019-lee}. 

We represent an exercise motion with sequential joint coordinates from a Kinect v2 sensor and extract various kinematic features similar to \cite{Lee2019LAQ}: joint angles (e.g. elbow flexion, shoulder flexion, elbow extension, shoulder abduction, the tilted angle of head, spine, and shoulder), speed related features (e.g. speed, acceleration, jerk, etc.) on wrist and elbow joints, and normalized relative trajectory (i.e. the Euclidean distance head and wrist and head and elbow). 

For feature selection, this paper utilizes reinforcement learning (RL) to dynamically identify salient features of assessment for individual patient's motions \cite{lee2020iha}. Specifically, we apply Double Q-learning \cite{van2016deep} to train an agent that sequentially decides whether another feature is necessary to assess an exercise while receiving a negative reward for requesting an additional feature or misclassification. Although the classical approaches of feature selection (e.g. filter, wrapper, embedded methods) select a fixed set of features with training data for all patients, our approach of feature selection with RL finds an optimal set of features for individual patient's motions \cite{lee2020iha}. Thus, we hypothesize and demonstrate that feature selection with RL is beneficial over classical feature selection approaches to generate personalized rehabilitation assessment and feedback. 

For a machine learning (ML) model to predict assessment, we utilize a Neural Network (NN) while grid-searching various architectures (i.e. one to three layers with 32, 64, 128, 256, 512 hidden units) and learning rates (i.e. $0.0001, 0.005, 0.001, 0.01, 0.1$) using \textit{`PyTorch'} libraries \cite{paszke2017automatic}. \textit{`ReLu'} activation and \textit{`AdamOptimizer'} are applied, and a model is trained until the tolerance of optimization is $0.0001$ or the maximum $200$ iterations. 

\section{Dataset of Three Upper-Limb Exercises}
For the evaluation, this paper utilizes the dataset of three task-oriented, upper-limb stroke rehabilitation exercises suggested by therapists: \textit{`Bring a cup to the mouth', `Switch a light on'}, and \textit{`Move forward a cane'} \cite{lee2020opportunities}. Fifteen stroke patients with different levels of functional abilities (37 $\pm$ 21 Fugl Meyer Scores) and 11 healthy subjects participated in data collection with a Kinect v2 sensor (Microsoft, Redmond, USA) that records the trajectory of body joints at 30 Hz. A patient performed 10 repetitions of each exercise with both patient's affected and unaffected sides. A healthy subject performed 15 repetitions of each exercise with the subject's dominant side. 
Two therapists annotated the dataset to implement our approach and compute therapist's agreement (TP in Table \ref{tab:results-avg}). 




\section{Results}
The machine learning (ML) model with reinforcement learning based feature selection (ML - RL) achieves the good agreement level with therapist's annotation: 0.7973 - 0.8331 average F1-scores on three exercises, which is comparable with therapist's agreement (TP in Table \ref{tab:results-avg}). In addition, our approach (ML - RL) achieves 0.11 higher average F1-score than the ML model with Recursive Feature Elimination, one of classical feature selection methods (ML - RFE). Thus, these results show that our approach can perform better to generate personalized assessment and corrective feedback. 
For the interaction with patients, we implement the interface that presents the tracked joints of a patient's exercise motion, predicted assessment on patient's exercise performance, and real-time audio and visual corrective feedback (Figure \ref{fig:interface}). 

\begin{table}[htp]
\centering
\caption{Performance (F1-scores) of our approach (ML - RL), the baseline approach with Recursive Feature Selection (ML - RFE), and therapist's agreement (TP)}
\resizebox{\columnwidth}{!}{%
\begin{tabular}{ccccc} \toprule
 & Exercise 1 (E1) & Exercise 2 (E2) & Exercise 3 (E3) & Overall \\ \midrule  \midrule 
  ML - RL & 0.8331$\pm$	0.0059 & 0.7973 $\pm$ 0.0867
 & 0.8053 $\pm$	0.0496 & 0.8119 $\pm$ 0.0526\\ \midrule
 ML - RFE & 0.6742 $\pm$ 0.0715 & 0.7628 $\pm$ 0.1708
 & 0.6415 $\pm$	0.0806 & 0.6928 $\pm$ 0.1147 \\ \midrule
 \begin{tabular}[c]{@{}c@{}}TP\end{tabular} & {0.7455 $\pm$ 0.2054} & {0.8147 $\pm$ 0.1522} & {0.7254 $\pm$ 0.1838} & {0.7619 $\pm$ 0.1626} \\ \bottomrule
\end{tabular} 
}
\label{tab:results-avg}
\end{table}



\section{Conclusion and Future Work}
This paper describes an interactive approach of a socially assistive robot that applies reinforcement learning for dynamic feature selection on individual patient's rehabilitation exercises to assess the quality of motion and generate personalized corrective feedback. 
The evaluation with the annotated dataset of three stroke rehabilitation exercises shows that our approach achieves good congruence with therapist's annotation, but also allows to generate transparent and personalized corrective feedback. In future, we will evaluate the usefulness of personalized corrective feedback from the NAO robot (Figure \ref{fig:setup}) for coaching post-stroke subject's rehabilitation exercises. 

\addtolength{\textheight}{-12cm}   

\bibliographystyle{IEEEtran}
\bibliography{main}

\end{document}